# Digital Identity: The Effect of Trust and Reputation Information on User Judgement in the Sharing Economy


Mircea Zloteanu[1*], Nigel Harvey[2], David Tuckett[3], and Giacomo Livan[1,4]

University College London

[1] Department of Computer Science, Faculty of Engineering, University College London, London, UK

[2] Department of Experimental Psychology, Division of Psychology and Language Sciences, University College London, London, UK

[3] Centre for the Study of Decision-Making Uncertainty, University College London, London, UK

[4] Systemic Risk Centre, London School of Economics and Political Sciences, London, UK

* m.zloteanu@ucl.ac.uk (MZ)


Word count: 6,065



# Abstract


The Sharing Economy (SE) is a growing ecosystem focusing on peer-to-peer enterprise. In the SE the information available to assist individuals (users) in making decisions focuses predominantly on community generated trust and reputation information. However, how such information impacts user judgement is still being understood. To explore such effects, we constructed an artificial SE accommodation platform where we varied the elements related to hosts' digital identity, measuring users' perceptions and decisions to interact. Across three studies, we find that trust and reputation information increases not only the users' perceived trustworthiness, credibility, and sociability of hosts, but also the propensity to rent a private room in their home. This effect is seen when providing users both with complete profiles and profiles with partial user-selected information. Closer investigations reveal that three elements relating to the host's digital identity are sufficient to produce such positive perceptions and increased rental decisions, regardless of which three elements are presented. Our findings have relevant implications for human judgment and privacy in the SE, and question its current culture of ever increasing information-sharing.

# Introduction

## Background

The Sharing Economy (SE) describes a growing ecosystem of online platforms devoted to the exchange of goods and services [1]. While a precise and encompassing definition of "sharing economy" is still debated within academia and business [2], the concept is grounded in peer-to-peer (P2P) enterprise, providing individuals with temporary access to the resources of other individuals [3], while using the platforms simply as an exchange mediator [4,5].

The SE has seen fantastic growth and high adoption rates among the general population [5], creating value on an unprecedented scale [4,6], and rivaling well established sectors of the traditional economy [7]. The types of platforms that have emerged around the SE paradigm range across all domains and markets, from accommodation (e.g., Airbnb) and taxi services (e.g., Uber) to household appliances (e.g., Zilok) and clothes (e.g., GirlMeetsDress).

## Trust and Reputation

In most marketplaces, individuals are aware that the services they request are subject to regulations, consumer protection laws, and monitoring by governmental bodies, ensuring a degree of liability, security, and safety. Within the SE, such protections are reduced [2], as SE platforms offer direct and largely unmediated interaction between individuals who have never met before neither offline nor online [1,4]. In this respect, SE operations require a high level of trust on behalf of all parties involved [8], which needs to be established in the absence of indicators typically employed to signal quality or reliability in traditional markets.

In recent years, the role of trust in the online environment has received



significant attention [9–15]. While its precise definitions may vary depending on the context, here trust refers to the psychological state reflecting the willingness of an actor to place themselves in a vulnerable situation with respect to the actions/intentions of another actor, in the absence of a direct ability to monitor or control the other party [16,17]. Thus, trust always involves some aspect of risk, uncertainty, vulnerability, and the expectation of reciprocation.

The ability to infer the trustworthiness of the individuals with whom users wish to engage is fundamental to the SE's operation [8,18–22]. This process is typically facilitated by requiring SE users to establish a digital identity (DI; namely, what the user decides to share on the platform and the information generated by the community).

Individuals in the SE (i.e. users), themselves, gravitate towards information relating to the trustworthiness of the individuals they wish to engage with before deciding to interact [23–25]. Furthermore, research has shown that trust, in turn, has a significant impact on user decision-making and behavior [9,11,13,26].

Another component driving user behavior within the SE is user reputation. Reputation in the SE reflects the perception the community has towards a user, through their longevity on a platform, contributions to the community, and outcome of past engagements (typically, represented as numeric or valence scores). Reputation can be considered an aggregate representation of trust towards a certain individual or entity. The combination of these factors is conducive to trust towards users on the platform [25]. Thus, in the SE a good reputation is considered as an alternative to trust [21]. Hence, trust and reputation can be considered the most valuable commodities within the SE [4,27,28].

Trust and reputation information are relevant due to the implicit information asymmetry and economic risks of SE platforms, forcing users to rely on such elements



to inform their decision-making [23,27,29,30]. The success of such a market relies on its ability to reduce uncertainty in its P2P interactions [4]. In such markets, often, the only source that allow people to infer the credibility of another user is to relying on their digital identify information. Thus, trust is crucial for turning a user's uncertainty into a definitive request to use a service [18].

## Digital Identity

At the core of any SE platform are systems that provide *reputation* building information [31], which typically aggregate *subjective* user generated content (UGC) into a reputation score; this in turn forms the core of a user's DI on the platform. Most SE platforms actively promote mechanisms through which users can *share* information, *rate* others, and build a *reputation* on the platform. Such content usually takes the form of numerical (e.g., ratings between 1 and 5) and text reviews. Each user has to carefully consider the reputation they foster within the community, ensuring that their identity convinces others that they are trustworthy and would want to interact with them.

Depending on the platform, the above information is typically complemented with more *objective* data through which users can signal trust and build relationships [32]. These include, but are not limited to, identity verification, photos of sellers and their goods, and sellers' contact information [33]. The combination of these two types of systems acts to reduce the implicit uncertainty of operating in such markets [16,19,34]. Yet, still little is known about how SE users incorporate such information into their decision-making processes.

## Aims & Research Questions

The central research question we address is whether SE users are able to integrate the wealth of available information on their peers' profiles when deciding with



whom to interact. We predicted that SE users are strongly affected by the presence of trust and reputation information (TRI), leading to changes in their judgements towards hosts and the services they offer, possibly without even realizing the strength of such effects.

We tested this prediction by designing an artificial accommodation platform, which allowed us complete control over the type and amount of information displayed to users. The profiles were generated to resemble accommodation SE platforms (e.g., Airbnb). Over three experiments, individuals (users) were presented with profiles reflecting hosts wishing to rent a private room in their house. The aim was to understand whether the TRI available on hosts' profiles has an effect on users' decisions to rent a private room advertised by a host. In Study 1, the effect of providing users with different amounts of host related TRI was assessed. This contrasted host profiles containing minimal information (akin to those of new users on a platform), with profiles containing the full information typically seen on such sites, and with profiles where the information presented was user-selected (allowing for some understanding of users' underlying thought processes). Study 2 addressed the latter point specifically, focusing on the importance of selecting versus being given the information of interest. Finally, in Study 3, the importance of the type of information users see was addressed. Here, the aim was to uncover whether specific TRI elements impact user judgements.

# Study 1

The first study investigated the effect of providing users with different amounts of host-related TRI, over three conditions: Hidden, Visible, and Reveal. This contrasted host profiles lacking TRI (Hidden), with full profiles (Visible), and profiles where only partial user-selected information was presented (Reveal). The effect of differences in



the amount of TRI on user judgments was measured on ratings of host credibility, trust, sociability, rent decisions, and confidence. It was predicted that the increased amount of information available would impact users' perceptions of hosts, resulting in differences in ratings, and decision to rent the private rooms. Additionally, allowing users to select host TRI would reveal which elements facilitate decision-making on SE platforms.

# Methods

## Participants and Design

A total of 160 participants were recruited online through Amazon's Mechanical Turk (mTurk; www.mturk.com) in exchange for $1.00. After deleting incomplete cases ($n$ = 36) the final data encompassed 124 participants (65 males, 58 females, one undisclosed; $M_{Age}$ = 35.11, $SD$ = 10.52; range: 20-73). Informed consent was obtained from all participants. This study has been ethically reviewed and received ethics clearance from the University College London Research Ethics Committee (CEHP/2015/534).

An independent-samples design was used, with three levels of Profile (Hidden, Visible, and Reveal). Participants were measured on multiple dependent variables: rent decision, confidence in decision, perceived sociability of host, trustworthiness of the host, and credibility of information (also, see **S2 Additional Dependent Variables**).

## Stimuli

Typical SE accommodation profiles were created specifically for the purposes of this experiment, using the Gorilla platform (www.gorilla.sc). These contained the elements generally featured on such sites, with the addition of two elements. The profiles were described as representing a "private room" in the host's house that they



wished to rent out to potential guests (see Fig.1).

--- Approximate Position of Fig 1. ---

The elements were as follows: a photo of the advertised room, a description of the room, a photo of the host, host verification, two guest reviews, two host reviews, online market reputation, social media presence, number of reviews, and star rating (for examples, see **S1**).

Certain factors were controlled in the creation of the profiles. To ensure quality and ecological validity, all elements were created to reflect the general ratings and content observed on SE platform profiles [35,36]. Two new elements were uniquely created to introduce further digital identity cues about the hosts: "social media presence" and "online market reputation". This decision stems from our perception of SE platform trends towards increased information-sharing and the role that users' DI has on their online perception [24]. The focus was on how cross-platform user reputation impacts decision-making on other platforms (i.e. the reputation of the user on one P2P platform influencing how they are perceived on another P2P platform). For a comprehensive description of element creation and controls implemented see **S1**.

## Procedure

Participants were randomly assigned to one of the three Profile conditions; Hidden ($n = 42$), Visible ($n = 40$), and Reveal ($n = 42$). They were first given a series of pre-task questions, including demographics (age, gender, and ethnicity), and SE usage (see **S2 Pre-task**). They then received instructions specific to their condition and were provided with an example profile, familiarizing them with the layout of the profiles, the type of information available, and the responses they would need to provide.



During the main task, users saw one profile at a time, and were asked if they wanted to "Rent" or "Not Rent" (binary, forced-choice). The other DV questions were measured on a 10-point Likert-type scale. Confidence was phrased as "How confident are you in your decision to Rent/Not rent?" (1 - *Not at all Confident* to 10 - *Very Confident*). Sociability was phrased as "How sociable do you think the host would be?" (1 - *Not at all Sociable* to 10 - *Very Sociable*). Trustworthiness was phrased as "How would you rate the trustworthiness of the host?" (1 - *Very Untrustworthy* to 10 - *Very Trustworthy*). While credibility was phrased as "How credible do you think the information about this room is?" (1 - *Not at all Credible* to 10 - *Very Credible*).

This was repeated over 10 trials. After each trial they were given a "validation" question, to ensure they were paying attention to the information in the profile. This was in the form of a question about a specific element on the profile (e.g., "What were the colors of the walls?", with an open-ended response). The elements comprising each profile were randomized between participants, reducing the artificial influence of specific combination on the responses.

In the Hidden condition, users saw a host's profile with minimal information presented about a room offered. This was limited to a photo of the room, a picture of the host, and a description relating to the room; these were always shown regardless of the profile condition. In the Visible condition, users saw a fully populated host profile, containing all the elements detailed above. In the Reveal condition, for each profile users had three tokens to "spend" on revealing any information they desired to help in their judgements. The elements available were: host verification, guest reviews, host reviews, online market reputation, social media presence, number of reviews, and star rating. The information regarding the spending of the three tokens was recorded in each trial. After the main task was completed, participants had to answer several post-task



questions, and were debriefed (see **S2 Post-task**).

# Results

Users' ratings were summed across the 10 trials and analyzed on each of the five dependent measures based on the three Profile conditions.

For decisions to rent the rooms, a one-way analysis of variance (ANOVA) revealed a significant main effect of Profile, $F(2,121) = 3.44$, $p = .035$, $\eta^2 = .054$. Post-hoc Tukey's HSD tests showed a significantly higher number of rent decisions in the Reveal condition ($M = 8.05$, $SD = 2.23$) than the Hidden condition ($M = 6.60$, $SD = 2.43$) and, $p = .027$. No other comparisons were significant.

Confidence in rent decisions was not found to be affected by Profile condition, $F < 1$, $p = .893$, suggesting that the type and amount of information participants saw on the profiles did not influence their confidence. Overall, the average confidence was very high for rent decisions ($M = 7.58$, $SD = 1.3$ per profile).

Ratings of host sociability were significantly affected by Profile condition, $F(2,121) = 6.41$, $p = .002$, $\eta^2 = .096$. Post-hoc tests showed significant lower user ratings for the Hidden condition ($M = 60.81$, $SD = 16.07$) compared to both the Visible ($M = 70.83$, $SD = 13.69$), $p = .006$, and the Reveal conditions ($M = 70.50$, $SD = 13.56$), $p = .008$.

Similarly, a main effect of host trustworthiness based on Profile was found, $F(2,121) = 8.94$, $p < .001$, $\eta^2 = .129$. Post-hoc tests revealed significantly lower ratings in the Hidden condition ($M = 63.50$, $SD = 15.50$) compared to both the Visible ($M = 75.13$, $SD = 14.29$), $p = .001$, and the Reveal ($M = 74.36$, $SD = 12.07$) conditions, $p = .002$. No other significant comparisons were found.

Finally, a main effect of perceived credibility was found, $F(2,121) = 7.00$, $p <$



.001 , $\eta^2$ = .104. Post-hoc comparisons revealed that the Hidden condition ($M$ = 67.48, $SD$ = 14.64) produced significantly lower ratings than both the Reveal ($M$ = 76.69, $SD$ = 11.87), $p$ = .005, and Visible ($M$ = 77.05, $SD$ = 13.04), $p$ = .004, conditions.

As the data suggest a lack of difference in user ratings between the Visible and Reveal condition, Bayesian independent-samples t-tests were conducted to complement the frequentist analysis. Considering the null hypothesis of no difference between the two conditions, tentative results were found in support of this claim. For all measures, the Bayes factor equaled between $BF_{01}$ = 2.6 – 4.34, indicating that the data was around 3 to 4 times more likely under then null than the alternative hypothesis (**S3 Bayesian Analysis**).

## Triplet Analysis

To obtain a better understanding of user selection preference for elements in SE platforms, an analysis of user token "spending" patterns in the Reveal condition was conducted. Each trial from each participant was treated as a unique vector of element selection from the total seven available items. Thus, the triplets of each of the 42 participants over the 10 trials were analyzed, 420 triplets in total. The triplet selection patterns were compared to a null model of random selection, using a binomial distribution (for details, see **S4**).

Initially, with respect to the frequency of observation of certain elements, a few results are relevant. It was found that 89.5% of the selected triplets featured *at least* one of the two following elements: "star ratings" and "guest reviews". Furthermore, 47.1% of the triplets featured *both* items, suggesting a strong user preference towards the two items. When testing for the over-representation of triplets, we find three combinations to be statistically significant at the 1% univariate level. These are: "star ratings + guest reviews + number of reviews", "star ratings + guest reviews + host verification", or



"star ratings + guest reviews + host reviews". These were interpreted as the triplets of TRI information users prefer most to aid in their decision-making.

# Discussion

This study finds that providing users with typical SE platform TRI increases their positivity towards their peers, rating them higher on sociability, trustworthiness, and credibility. Importantly, this also resulted in an increased number of rent decisions. Differences were observed when comparing user ratings in the Hidden condition, where all but the basic information was present, with the Visible and Reveal condition, where reputation and trust information was provided to varying degrees. The results also suggest that seeing a full profile (Visible) or one with only partial, but user-selected, information (Reveal) results, on average, in similar judgements towards hosts. Considering the difference in information between the two conditions (i.e. seven elements in the Visible condition and three user-selected elements in the Reveal condition), two explanations are proposed.

Potentially, the act of selecting which elements to see impacted users' perception, leading to the belief that the these elements are the most relevant/useful for their decision [37]. Thus, the act of selecting specific information may be generating this "positivity effect" (i.e. increased ratings towards hosts on all measures).

Alternatively, it may be that users rely only on a few elements when making their decisions, as argued by past decision-making research [37–39]. Thus, users may not be able to incorporate in their judgment the additional information present in a full profile.

A final noteworthy finding is that confidence in user decision was generally high and was not affected by the Profile condition. This suggests that the TRI is not a factor



that impacts how confident users are in their SE rental decisions.

# Study 2

Study 2 was devised to unpack the effects relating to the act of selection and the amount of information on user judgement, as the data indicated that users relying on three selected elements rated hosts similarly to participants which had access to all seven elements.

People often report using complex strategies and a large number of cues when making their judgements; however, empirical studies show that they fail to accurately use more than two or three cues at any one time, and incorrectly predict which ones play a role in their decision-making process [37,39–42]. Thus, the lack of difference due to the amount of information may be explained either by the fact that (1) users considered the elements they selected as being more informative towards their final decision, leading to equally positive judgements, or (2) users seeing more than three elements could not incorporate these into their decision-making (i.e. discounting information), thus producing no differences in judgement.

A modification of Study 1's design was implemented. Using the results from the Reveal condition, we selected element triplets to present to users directly and compared their responses to users that selected three elements. This ensured that the amount of information was kept constant (i.e. three pieces of TRI per profile), and that any difference in renting decision or perceptions of hosts would be a result of the act of selecting information.



# Methods

## Participants and Design

120 participants were recruited through mTurk, in exchange for \$1.00. After deleting incomplete cases ($n = 3$), 117 participants remained (65 males, 52 females; $M_{Age} = 36.21$, $SD = 10.35$; range: 20-69). The design was similar to that of Study 1, with Profile having two conditions: 3-Seen, 3-Reveal. Participants were measured on the same DVs as Study 1.

## Stimuli

The profiles in the 3-Seen condition contained only three TRI elements to assist in user decision-making, selected based on the data from the previous study. In this condition each profile contained one of the three triplet combinations whose selection frequency was found to be significant in Study 1: "stars + guest reviews + number of reviews", "stars + guest reviews + host verification", or "stars + guest reviews + host reviews". In the 3-Reveal condition, the same profile generation procedure as in Study 1's Reveal condition was employed. This allowed for a direct comparison of the effect of choice on user judgement, as information was kept constant in all conditions.

## Procedure

The procedure was identical to that of Study 1, but here participants were randomly assigned to one of the two Profile conditions: 3-Seen ($n = 61$) or 3-Reveal ($n = 56$). Within the 3-Seen condition, participants were further randomly divided into the three sub-conditions, based on the triplet they saw. Thus, each participant in the 3-Seen sub-condition saw a specific triplet throughout his/her trials (randomized between participants with a rate of 1/3). In all conditions participants saw 10 profiles and



responded to questions regarding each profile using the same scales as Study 1.

# Results

Multiple independent samples t-tests were conducted to compare the responses in the 3-Seen and 3-Reveal conditions for each DV. The data revealed no significant differences in any of the responses users gave based on Profile condition, on any of the measured variables, $ts \leq 1$, $ps > .05$. This supports the explanation that users do not rely on more than three pieces of information judging a host's profile (for additional analyses, see **S5**).

To provide more support, Bayesian independent-samples t-tests were conducted on the data, which did favor this conclusion. Moderate to substantial support in favor of accepting the null of no difference between the 3-Seen and 3-Reveal conditions was found, $BF_{01}$ between 2.29 and 3.77 for all DVs (**S5 Bayesian Analysis**).

# Discussion

The data suggests that users are not affected by the act of selecting which TRI they want to see before making a rental decision beyond simply being presented three pieces of information. On all measured variables, users provided similar responses in the 3-Seen and 3-Reveal conditions. This suggests that users' judgements are influenced by seeing at least three elements of TRI, but are unaffected by either the act of selecting said information or seeing more than three (**S5 Comparison to Study 1**).

# Study 3

People find uncertainty unpleasant and seek to reduce it [43]. Relying on information provided by others may assist with this issue, allowing users to reduce the



number of alternatives and task complexity [44]. However, the results from the above studies raise questions about the actual usefulness to users of the information available on SE platforms.

Is the information in the elements presented to users impacting decisions, or do users simply need three elements to believe they are making an "informed" decision? To understand this, the third study assessed whether *specific* triplets result in differences in users' judgements about hosts' credibility, sociability, trustworthiness, and decision to rent a room. Here, users saw one of three options. From Study 1's Reveal condition, specific triplets were chosen to reflect the least selected triplet combinations (3-Avoided), the three most selected triplet combinations (3-Wanted), and triplets containing a random combination of the remaining elements (3-Random).

A significant difference in users' judgements between conditions would support the claim that users have a preference for what information they use when making their rental decisions, which in turn affect their ratings of hosts. Alternatively, a lack of a difference would suggest that users simply require the presence of three elements for the positivity effect to occur.

# Methods

## Participants and Design

189 participants (103 males, 86 females; $M_{Age}$ = 34.09, $SD$ = 10.23; range: 19-48) were recruited through mTurk, in exchange for $1.00. The design was similar to the previous studies, with Profile having three conditions: 3-Wanted, 3-Avoided, 3-Random. Participants were measured on the same DVs as the previous studies.

## Stimuli

In the 3-Wanted condition, profiles were generated using the same three triplet



combinations as in Study 2's 3-Seen condition. In the 3-Avoided condition, profiles were comprised of three triplet combinations chosen so as to be the least selected combinations by users in Study 1's Reveal condition (see **S4**). In these sub-conditions, "host verification" and "host reviews" were always shown together, while the third element was either "social media presence", "online market reputation", or "number of reviews". In the 3-Random condition, profiles were comprised of randomly selected triplet combinations formed from the remaining unused combinations.

## Procedure

Participants were randomly allocated to one of the three Profile conditions: 3-Wanted ($n = 67$), 3-Avoided ($n = 67$), 3-Random ($n = 55$). For the 3-Wanted and 3-Avoided conditions, participants were further divided into their respective three sub-conditions. Participants saw 10 host profiles in each condition containing a consistent triplet of information throughout. They responded to each profile using the same scales as in the previous studies.

# Results

Multiple one-way ANOVAs were conducted to uncover any differences in users' judgments based on Profile condition. The data did not reveal any statistically significant effects of Profile on any of the measured variables, $F$s $< 1$, $p$s $> .44$ (for additional analyses, see **S6**).

To provide additional support for the findings of no differences in user judgments based on specific elements, the frequentist analyses were complemented with a Bayesian approach. Testing the assumption of no difference from the null hypothesis, the Bayesian ANOVA conducted on rent decisions between the three Profile conditions revealed a Bayes factor of $BF_{01} = 9.05$, indicating that the data was around 9 times more



likely under the null than the alternative hypothesis. For confidence, a $BF_{01} = 13.35$ was found in favor of the null. With regards to host ratings, for sociability a $BF_{01} = 14.36$ was found, for trustworthiness a $BF_{01} = 14.01$ was found, and for credibility a $BF_{01} = 12.15$ was found.

## Discussion

The data consistently shows that the information contained in the individual TRI elements does not influence judgment beyond their simple presence. No differences in user judgements were found when comparing triplets that users selected most often (3-Wanted), or the ones they avoided selecting (3-Avoided), or simply showing three random elements (3-Random). The findings suggest that providing users with at least three TRI elements to aid their decision-making is sufficient to produce a strong positivity in their judgements of hosts on SE accommodation-style platforms.

# General Discussion

The SE's proliferation brings opportunities for individuals, but also potential, and yet unknown, risks. The information individuals rely on in SE platforms differs significantly from what consumers generally rely on in traditional marketplaces. Namely, there is an overreliance on UGC to assess the trustworthiness and reputation of other users. In this respect, the novelty of the SE and its multiple platforms raises many questions regarding people's judgments and their ability to use such information to make informed decisions.

To understand the impact of trust and reputation based systems on human judgment, the current paper investigated the role of DI information on user judgments in an artificial SE accommodation platform. Over three studies, the data showed that users are strongly influenced by the presence of TRI, resulting in an overall positivity towards



hosts and increased tendency to rent rooms. The ratings users gave to hosts on measures of sociability, credibility, and trustworthiness increased significantly when they were provided information relating to the hosts' DI, compared to seeing a profile lacking such information. However, the amount or specificity of this information did not seem to impact judgement.

The findings of Study 1 speak to the effects that P2P platforms have on people's decision-making process. In line with past research, TRI is found to have a significant impact on users' decision to engage with others on P2P platforms, and to seek their services [24]. Here, we find that even when maintaining the quality of the service or product constant (i.e. the rooms), users perceive these in a more positive light if they are provided with TRI regarding the provider (i.e. host).

The sharp contrast in rent decisions between the Hidden and other profile conditions would suggest that TRI is used in deciding not only our perceptions of other peers on SE platforms, but also of the likelihood we will use their goods or services. Viewing the Hidden condition profiles as reflecting new users on a platform, illustrates that even profiles devoid of TRI can have success (as seen by the overall high rent decisions). However, it is clear that profiles containing platform- and community-generated information have a significant advantage [25,31,34]. Furthermore, the lack of a statistically significant difference between the Visible and Reveal conditions implied either that users were discounting the extra information, resulting in no added positivity beyond that seen in the Reveal condition, or that the act of selecting which information to reveal compensated for the information difference.

Study 2 examined these explanations. Here, the data suggested that simply seeing three elements is sufficient to generate an overall positivity towards hosts and to increase rental decisions, while the act of self-selecting information has no effect on



judgment.

Study 3 investigated whether specific information mattered towards this positivity; that is, whether different elements result in a different perceptions of hosts. Overall, the data strongly showed that specific combinations did not matter. Irrespective of the frequency with which users tend to select them naturally, presenting three TRI elements is sufficient to positively impact judgments and increase rent decisions.

The above findings resonate with the judgment and decision-making literature, expanding it to online user behavior. From this literature, we know that people are cognitive misers [45], rarely able to incorporate diverse information from multiple sources into their judgements. People are also subject to several biases that can influence their perception and the trust they place in others, as well as the risks they are willing to take. Rather than relying on rational and strategic process to make decisions, people tend to rely on quick and automatic rules-of-thumb to make their judgments [46]. Moreover, people are poor at estimating their own preferences for information [41], and are limited to around three cues when making judgments [42].

Thus, while people show a strong selection preference for specific TRI elements in SE environments, these may not reflect specific differences in how the information is perceived or used by individuals. Indeed, even in novel environments, where people have minimal information on how different information affects outcome, they seem to develop strong preferences over time (see coherent arbitrariness [47]). They may, nonetheless, prefer TRI due to social convention (i.e. they use it because others seem to use it [48] or because they are accustomed to relying on it [49]).

Lastly, the literature argues that TRI serves to reduce the uncertainty experienced in SE environments [4,34]. However, across all studies users' decision confidence was found to be stable and overall high, regardless of the amount or type of



TRI presented. The data favors our interpretation that TRI produces a "positivity effect" on user judgement, rather than an "uncertainty-reduction effect" (cf. [50]).

# Implications

Decision-making in an online P2P environment can be a complex task. This process is compounded by issues with the information provided to users to aid in their decision-making that exist in the SE, two central ones being the overall positivity of such information and, consequently, its low diagnosticity [21,36]. Indeed, ratings on SE platforms show a stronger bias towards high ratings than on other P2P platforms [35,36], which severely reduces their usefulness to users.

Despite this, there is a trend for increased UGC on online platforms. Users seem more willing to provide such information, even when private and potentially identifying, and platforms themselves are incentivizing this type of information-sharing [51]. But, our data show that more of such information may not assist people in any meaningful way, which in turn suggests that both platforms and their users gain no benefit from collecting and sharing more information than is currently available. However, these considerations do not necessarily imply that limiting the proliferation of UGC would have no consequence.

First, research finds that users trust UGC more than objective metrics when making their decisions, and carry more weight in the decision-making process than other forms of information [24,52]. This is supported by the current data, finding that users show a preference for selecting elements that result from the aggregate ratings or other users' testimonials (e.g., guest reviews), compared to platform-generated information (e.g., host verification).

Second, attempting to reduce the amount of TRI from existing or emerging platforms may backfire in terms of user perception. Users may *expect* specific



information to be present (even if not used), with its absence leading to more negative

appraisals or to avoidance [53].

Past research has argued that reputation systems and user-generated reviews

may have the primary purpose of allowing users to learn more about each other before

engaging in any interactions or transactions, acting as a monitoring and policing system

[2,54,55]. However, the current findings demonstrate that this information can also act

as a strong influencer in the perceptions of others, leading users to see peers in a more

positive light.

A consideration that emerges from the current findings relates to the concept of

online privacy and how people construct their digital identity. A culture of information-

sharing is forming, under the guise of more informed decision-making, which may force

individuals wanting to participate in these communities to share private and sensitive

information without any benefit to the community or individual decision-making

outcomes. The current data show it is unnecessary for users, beyond a certain point, to

provide such information, and may even prove detrimental in the long run. In this

respect, we advise caution in how SE platforms choose to expand and implement their

reputation-based systems.

## Future Directions

The aim of the current research was not to assess whether specific reputation and

trust elements are useful to discriminate among different options, but rather to

understand how the *presence* of such information affects user judgement and decision-

making in a setting that closely follows real-world patterns.

A natural extension of the current research is to understand how accurate

individuals are at classifying profiles based on their quality. The current design

considered the effect of cue diagnosticity to the extent that no one particular element



provided specific information to classify a room as "good" or "bad", but aimed to reflect the natural distribution seen on SE platforms [35,36]. However, if the profiles users saw varied more in terms of quality and uncertainty, would ratings reflect these differences or would they continue to show a positivity effect? Introducing variability and an element of diagnosticity into the information users see on such profiles may provide a more complete image of human behavior in the SE. Similarly, this can extend into considering the difference in effect and strength that negative information has on judgments compared to positive information [56]. This research is currently being undertaken.

## Conclusion

Currently, the effect of TRI on user behavior in the SE was investigated. The focus was on how presenting users with information about hosts' DI, in an accommodation SE platform, would impact their perceptions of hosts and the likelihood of renting their private rooms. Over three studies, the data consistently shows that users find hosts whose profiles display TRI as more trustworthy, credible, and sociable. More importantly, they also rent more properties if such information exists. Despite users showing a consistent and strong preference for specific information, they demonstrate this positivity in judgement from seeing any three elements relating to the hosts' DI. These findings illustrate how TRI can affect user decision-making, cautioning people on the risks of relying too heavily on this information. Research should carefully consider how information relating to trust and reputation on SE platforms can impact user judgement.

## Supporting Information

**S1 Profile Elements** (PDF)



**S2 Additional Measures** (PDF)

**S3 Study 1 Demographics and Supplementary Analyses** (PDF)

**S4 Triplet Analysis** (PDF)

**S5 Study 2 Demographics and Supplementary Analyses** (PDF)

**S6 Study 3 Demographics and Supplementary Analyses** (PDF)

# Acknowledgements

Giacomo Livan and Mircea Zloteanu acknowledge support from an EPSRC Early Career Fellowship in Digital Economy (Grant No. EP/N006062/1).

# Author Contributions

Conceived and designed the experiments: MZ GL NH DT. Performed the experiments: MZ. Analyzed the data: MZ GL. Contributed reagents/materials/analysis tools: MZ GL NH DT. Wrote the paper: MZ GL.

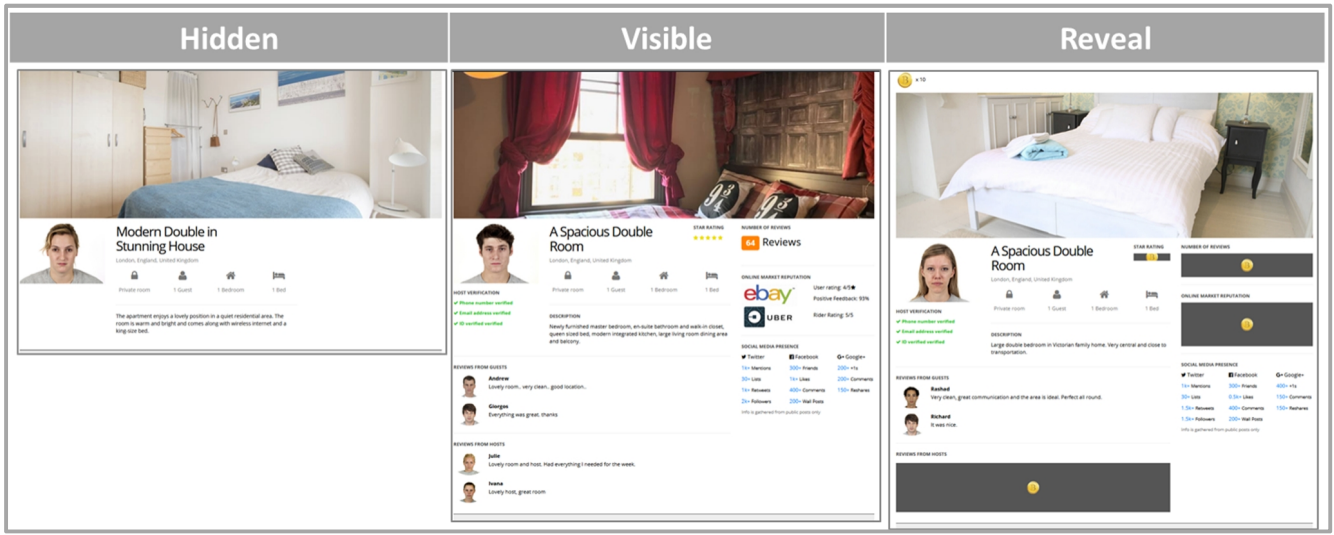

**Fig.1. Example of host Profiles in each condition, indicating which elements were visible to users.**

# S1 Profile Elements

Below is a description of each element comprising the profiles of hosts shown to participants, along with an example of how it appeared in the experiments.

**Room Photo**

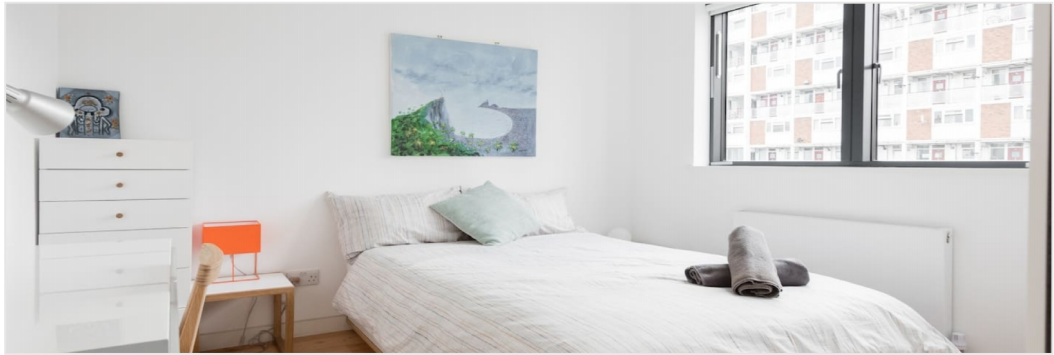

*Creation*: The room images were selected from Airbnb. These were limited to a single borough of London, within one standard deviation (SD) of the average price for a room in the area, equating quality and amenities offered.

**Title**

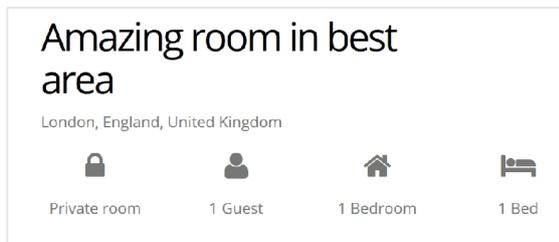

*Creation*: Selected directly from Airbnb. These were controlled for word length (3-7), location information, and other identifying cues.

**Description of room**

DESCRIPTION

A modern & clean private double room (private WC) with lots of natural light & a comfortable bed, within a very large 2-bedroom loft style apartment with a south-facing balcony where you can sit and enjoy an amazing view.

*Creation*: Room descriptions were taken from Airbnb. These focused on descriptive information, did not contain identifying location cues or opinions, and were constrained to between 20-50 words.

**Host Photo**

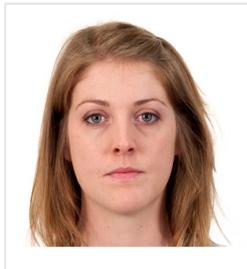

*Creation*: The profile photos were created from the Chicago Face Database, which offers descriptions of each photo with ratings on several factors [1]. Here, facial dominance, trustworthiness, attractiveness, and ethnicity were controlled, as differences in facial traits can influence decision-making (e.g., [2,3]. Only images that did not deviate more than ±1 SD from the mean on each characteristic, within one ethnicity (White-Caucasian), were selected. The same care was given to the "guest" and "host" reviews photos, without limiting ethnicity.

**Star Rating**

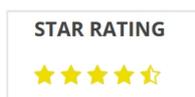

*Creation*: The star ratings on the profiles were dynamically and randomly generated for each profile. The stars ranged from 1 to 5 in half-star increments, however, the distribution was limited to 4 to 5 stars, as is typical of such profiles [4,5].

**Host Verification**

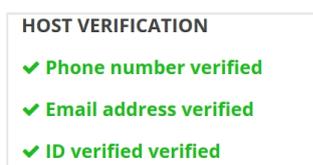

*Creation*: These represented the platform's verification of the host's identity. This always

showed a green tick-mark for all three verifications available: mobile, ID, and address.

## Number of Reviews

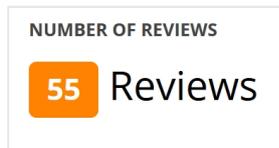

*Creation*: The number of reviews was displayed as a badge with a number inside. This represented the total number of reviews that existed for the specific room, acting as a metric for reputation and longevity. The number was randomly generated starting with 50 + a random number from 0-30.

## Guest Reviews

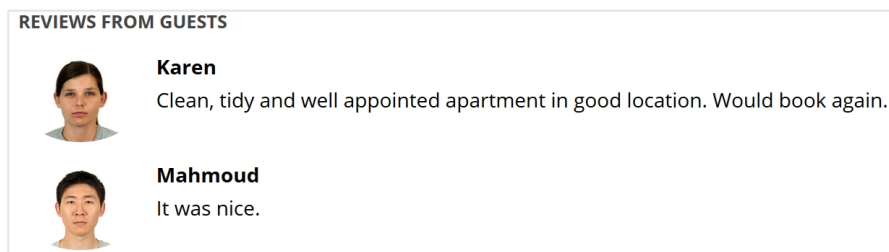

*Creation*: The guest and host reviews were taken directly from Airbnb. Guest reviews reflect the testimony of past users that have stayed with that particular host, while host reviews are testimonials from other hosts on the platform that have also stayed with that host in the past. The comments focused on the room quality, contained no identifying information, and did not exceed 15 words.

## Host Reviews

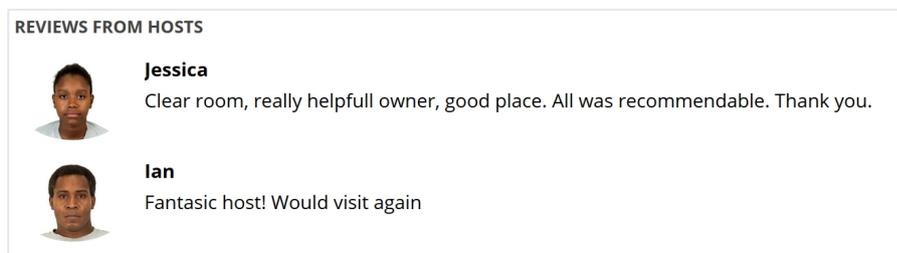

*Creation*: See Guest Reviews.

## Online Market Reputation

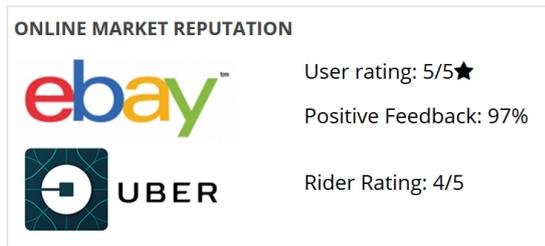

*Creation*: Social media presence was composed of the host's online presence on Twitter, Facebook, and Google+, generated randomly for each profile. For Twitter, the data included the number of "mentions" for the host (range 1 - 2k, with 1 decimal place (dp) increments), number of "lists" (range 10-30), number of "retweets" (range 1 – 2k, 1 dp), and number of "followers" (range 1 – 2k, 1dp). For Facebook, the data included number of "friends" (range 300 – 700), "likes" (range 0.5 – 1.5k, in 1 dp increments), "comments" (range 400 – 700), and "wall posts" (range 100 – 200). For Google+, the data was comprised of number of "+1's" (range 200 – 400), "comments" (range 100 – 200), and "reshares" (range 100 – 200).

## Social Media Presence



*Creation*: Online market reputation containing the host's user profile ratings on two other platforms, eBay and Uber. For eBay this represented the "user rating", as a randomly-generated fractional rating from 1 to 5 (effective range 4 – 5, with .5 increments; e.g., 4/5), and "positive feedback" with a percentage randomly generated (90 + random number between 0 – 10). For Uber the information contained the "rider rating", with a number ranging from 1 to 5, but a constrained range of 4 – 5 (.5 increments).

# S2 Additional Measures

# Pre-task questions:

Number of memberships on Sharing Economy sites/apps (e.g. Uber, Airbnb, TaskRabbit)

○ None
○ 1 - 2
○ 3 - 4
○ 5 - 10
○ >10

**Thinking about your usage and experience with sharing economy platforms (e.g., Airbnb, Uber), please answer the below questions regarding your behaviour and history on such sites/apps:**

How long have you been a member of or using such sites/apps?

| <1 month | 2 - 6 months | 8 - 10 months | 12 months | >3 years |
|---|---|---|---|---|

How many times have you used such sites/apps within the last year?

| 0 - 5 | 5 - 10 | 10 - 20 | 20 - 30 | >30 |
|---|---|---|---|---|

In general, how satisfied have you been with using such platforms?

| **Not at all Satisfied** | 1 | 2 | 3 | 4 | 5 | 6 | 7 | 8 | 9 | 10 | **Very Satisfied** |
|---|---|---|---|---|---|---|---|---|---|---|---|

How would you rate your sense of belonging to such peer-to-peer platforms?

| **None at all** | 1 | 2 | 3 | 4 | 5 | 6 | 7 | 8 | 9 | 10 | **Very Strong** |
|---|---|---|---|---|---|---|---|---|---|---|---|

**Please read the below statments, and rate how strongly you agree or disagree with them:**

1 - "Being accepted as a member of a group is more important than having autonomy and independence."

| **Strongly Disagree** | 1 | 2 | 3 | 4 | 5 | 6 | 7 | 8 | 9 | 10 | **Strongly Agree** |
|---|---|---|---|---|---|---|---|---|---|---|---|

2 - "Being accepted as a member of a group is more important than being independent."

| **Strongly Disagree** | 1 | 2 | 3 | 4 | 5 | 6 | 7 | 8 | 9 | 10 | **Strongly Agree** |
|---|---|---|---|---|---|---|---|---|---|---|---|

3 - "Group success is more important than individual success."

| **Strongly Disagree** | 1 | 2 | 3 | 4 | 5 | 6 | 7 | 8 | 9 | 10 | **Strongly Agree** |
|---|---|---|---|---|---|---|---|---|---|---|---|

4 - "Being loyal to a group is more important than individual gain."

| **Strongly Disagree** | 1 | 2 | 3 | 4 | 5 | 6 | 7 | 8 | 9 | 10 | **Strongly Agree** |
|---|---|---|---|---|---|---|---|---|---|---|---|

5 - "Individual rewards are not as important as group welfare."

| **Strongly Disagree** | 1 | 2 | 3 | 4 | 5 | 6 | 7 | 8 | 9 | 10 | **Strongly Agree** |
|---|---|---|---|---|---|---|---|---|---|---|---|

6 - "It is more important for a manager to encourage loyalty and a sense of duty in subordinates than it is to encourage individual initiative."

| **Strongly Disagree** | 1 | 2 | 3 | 4 | 5 | 6 | 7 | 8 | 9 | 10 | **Strongly Agree** |
|---|---|---|---|---|---|---|---|---|---|---|---|

## Post-task questions:

**Please answer to the best of your abilities.**

When usually making rental decision online, which information source carries the most weight in your decision to "Rent" or "Not rent"? **(select 3 max)**

| | Picture of the room |
| | Host's profile image |
| | Description of the room |
| | Profile's Star Rating |
| | Total number of Reviews |
| | Online Reputation Scores |
| | Social Media Presence information |
| | Reviews from other Hosts |
| | Reviews from other Guests |
| | My Intuition |
| | None of the above |

If you ticked "None of the above", please write down how you made your judgements:

To help in making such rental decisions, would you like to see more cross-platform information about the hosts? *(i.e. their reputation on other peer-to-peer platforms, or their social media activity information)*

○ Yes

○ No

○ Not sure

Please write down any other information you felt was missing from the profiles which may have helped in your decision making: *[optional]*

What do you think the aim of this experiment was?

## Additional Dependent Variables:

Cognitive trust and Affective trust towards the hosts were also measured during the three studies. However, preliminary analyses indicated strong overlap with the overall Trustworthiness measure. The addition of this variable in analyses or the results provides no further insight into user judgment in this specific paradigm, so were not presented in the main text. The questions used to assess these measures are as follows:

- **Cognitive Trust**: 'Given the information displayed on the webpage, I have no doubt about this host's competence'' (10-point Likert Scale)
- **Affective Trust**: 'If I shared my concerns about the property with this host, I know s/he would respond constructively'' (10-point Likert Scale)

# S3 Study 1 Demographics and Supplementary Analyses

### Participant Demographics

| Ethnicity | | | | | | |
|---|---|---|---|---|---|---|
| **Asian** | **Black** | **Latino** | **White** | **Multi-racial** | **Other** | **N** |
| 6 | 14 | 5 | 98 | 0 | 0 | 123 |

| Sharing Economy - Memberships | | | | | |
|---|---|---|---|---|---|
| **None** | **1-2** | **3-4** | **5-10** | **>10** | **N** |
| 19 | 82 | 19 | 2 | 1 | 123 |

| Sharing Economy – Usage Length | | | | | |
|---|---|---|---|---|---|
| **<1 month** | **2-6 months** | **8-10 months** | **12 months** | **>3 years** | **N** |
| 29 | 22 | 19 | 35 | 18 | 123 |

| Sharing Economy - Usage Frequency | | | | | |
|---|---|---|---|---|---|
| **0-5** | **5-10** | **10-20** | **20-30** | **>30** | **N** |
| 67 | 26 | 17 | 8 | 5 | 123 |

| Sharing Economy - Satisfaction | | | | | | | | | |
|---|---|---|---|---|---|---|---|---|---|
| **1** | **3** | **4** | **5** | **6** | **7** | **8** | **9** | **10** | **N** |
| 6 | 2 | 4 | 23 | 8 | 19 | 38 | 14 | 9 | 123 |

| Sharing Economy – Sense of Belonging | | | | | | | | | |
|---|---|---|---|---|---|---|---|---|---|
| **1** | **3** | **4** | **5** | **6** | **7** | **8** | **9** | **10** | **N** |
| 16 | 6 | 13 | 5 | 28 | 16 | 13 | 12 | 8 | 123 |

## Gender Manipulation Check

An initial analysis considering gender differences based on the Profile condition was conducted, however, this did not reveal any significant effects on any of the measured DVs, $p$s > .05.

## Bayesian Analysis

To add more support for the explanation that there are no differences between the Visible and Reveal conditions, the frequentist analysis was supplemented with a Bayesian analysis. Multiple independent-samples t-tests were conducted under the assumption of no difference between the two conditions. Below is a table detailing the exact values for each dependent variable measured.

| Bayesian Independent Samples T-Test | | |
|---|---|---|
| | $BF_{01}$ | error % |
| Rent | 2.611 | 0.022 |
| Confidence | 4.340 | 0.024 |
| Social | 4.327 | 0.024 |
| Trust | 4.219 | 0.024 |
| Credible | 4.316 | 0.024 |

# S4 Triplet analysis

We assess the statistical significance of the three elements selected by users in the Reveal condition by contrasting them with those that would have been observed under a null hypothesis of random selection (i.e. a pattern of random "token spending").

In order to assess the statistical significance of these findings, we assumed a null hypothesis of random selection, i.e. assuming that each trial results from the random selection of 3 out of the 7 available elements. This allows to assign a p-value to each triplet. Assuming a certain triplet has been selected x times, this reads,

$$P_u(x) = 1 - \sum_{i=0}^{x-1} \binom{n}{i} p^i (1-p)^{n-i} \,,$$

where n denotes the total number of trials (42 participants x 10 trials each = 420), and p denotes the probability of selecting any particular triplet at random out of the 7 available options, which reads

$$p = {1} \Big/ {\binom{7}{3}} \sim 0.0286.$$

The above equation for P(x) denotes the complementary cumulative function of the binomial distribution, and computes the probability of observing the triplet under consideration x *or more* times. Symmetrically, the above framework can be used to assess the possible under-representation of certain triplets in the choices made by users. Namely, the p-value

$$P_d(x) = \sum_{i=0}^{x} \binom{n}{i} p^i (1-p)^{n-i}$$

evaluates the probability of observing a certain triplet x or less times under a null hypothesis of random selection.

We test significance at a univariate significance level of 1%, and we perform a Bonferroni correction to the above p-values to take into account the multivariate nature of our test (420 trials). When testing for the over-representation of triplets, we find 3 combinations

to be statistically significant at the 1% univariate level. These are: "stars + guest reviews + number of reviews", "stars + guest reviews + host verification", or "stars + guest reviews + host reviews".

When instead testing for the under-representation of triplets, we find 5 combinations to be significant at the 1% univariate significance level, which correspond to the only combinations never selected in any trial. Three of these were found to feature both the "Host Verification" and "Host Reviews" items, with the third one being, respectively, "Social Media Presence", "Online Market Reputation", or "Number of Reviews". This motivated our selection of triplets for Study 2.

We also employ a similar methodology to assess whether participants progressively converge over a set of preferred choices over time. In order to do this, we compare the overlap between choices in consecutive trials and, again, compare it to a null hypothesis of random selection. In particular, for each trial we count the number of participants whose triplet shares at least two items with the triplet of the previous trial. In analogy with the above p-values, we compute the probability of observing c or more users whose triplet shares at least two items with their previous choice as

$$P_t(c) = 1 - \sum_{i=0}^{c-1} \binom{u}{i} \pi^i (1-\pi)^{c-i},$$

where u = 42 is the number of users and

$$\pi = \frac{1 + \binom{3}{2}\binom{4}{1}}{\binom{7}{3}} \sim 0.371$$

is the probability of selecting at least two of the 3 items selected in the previous triplet at random. In the plot below we show such p-values for each set of trials, and we also report the corresponding Bonferroni corrected significance level. As it can be seen, we observe a statistically significant overlap between consecutive trials already from trial 2, illustrating a strong tendency to select the same items from the very beginning with very little exploration of other possibilities.

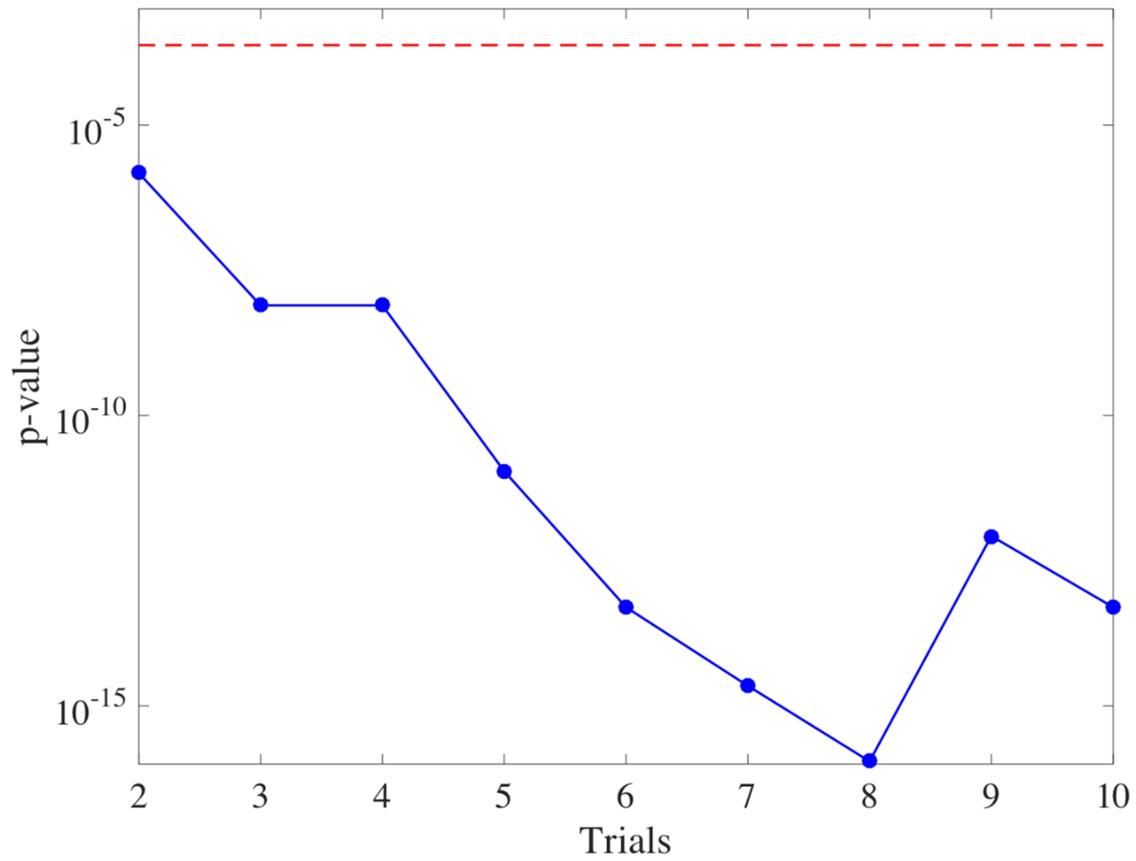

Fig 1. The plot illustrates the statistical significance of overlap in element selection between consecutive trials (i.e. if the selection of elements in one trial predicted the selection in the next trial). The x-axis denotes the overlap in selection between trial X and trial X-1. The y-axis represents the p-value of the overlap. The blue solid line shows the p-values for each trial comparison, while the red dashed line represents the Bonferroni corrected p-value level of significance.

# S5 Study 2 Demographics and Supplementary Analyses

## Participant Demographics

| Ethnicity | | | | | | |
|---|---|---|---|---|---|---|
| **Asian** | **Black** | **Latino** | **White** | **Multi-racial** | **Other** | **N** |
| 9 | 11 | 1 | 90 | 6 | 9 | 117 |

| Sharing Economy - Memberships | | | | | |
|---|---|---|---|---|---|
| **None** | **1-2** | **3-4** | **5-10** | **>10** | **N** |
| 10 | 82 | 20 | 5 | 0 | 117 |

| Sharing Economy – Usage Length | | | | | |
|---|---|---|---|---|---|
| **<1 month** | **2-6 months** | **8-10 months** | **12 months** | **>3 years** | **N** |
| 12 | 17 | 19 | 48 | 21 | 117 |

| Sharing Economy - Usage Frequency | | | | | |
|---|---|---|---|---|---|
| **0-5** | **5-10** | **10-20** | **20-30** | **>30** | **N** |
| 53 | 30 | 18 | 6 | 10 | 117 |

| Sharing Economy - Satisfaction | | | | | | | | | |
|---|---|---|---|---|---|---|---|---|---|
| **1** | **3** | **4** | **5** | **6** | **7** | **8** | **9** | **10** | **N** |
| 5 | 3 | 1 | 11 | 10 | 23 | 30 | 21 | 13 | 117 |

| Sharing Economy – Sense of Belonging | | | | | | | | | |
|---|---|---|---|---|---|---|---|---|---|
| **1** | **3** | **4** | **5** | **6** | **7** | **8** | **9** | **10** | **N** |
| 10 | 4 | 6 | 5 | 15 | 17 | 23 | 23 | 9 | 117 |

## Manipulation Check

Preliminary analyses revealed no statistically significant differences between the three sub-conditions of the 3-Seen condition, $p$s > .05; thus, all subsequent analyses collapsed across the sub-conditions.

Subsequently, an analysis considering gender differences based on the Profile conditions was conducted. This did not reveal any significant effects on any of the measured DVs, $p$s > .05.

Finally, to verify that the responses from Study 2's 3-Reveal condition ($n = 54$) did not differ from those of Study 1's Reveal condition ($n = 42$), the data of the two was compared on all measured DVs. In all cases, no significant differences were found between the two conditions, $t$s < 1, $p$s > .05.

## Bayesian Analysis

Independent-samples t-tests were conducted investigating the non-significant findings between the 3-Seen and 3-Reveal conditions uncovered when using the frequentist approach. These analyses were conducted on all measured DVs: rent decision, confidence in decision, trustworthiness, credibility, and sociability. As suggested, the data suggests that user behaviour did not differ if either allowed to select three TRI elements to aid their decision or simply be provided three elements.

| Bayesian Independent Samples T-Test | | |
|---|---|---|
| | $BF_{01}$ | error % |
| Rent | 2.288 | 0.017 |

| Bayesian Independent Samples T-Test | | |
|---|---|---|
| | $BF_{01}$ | error % |
| Confidence | 3.029 | 0.012 |
| Sociability | 3.768 | 0.007 |
| Trust | 2.034 | 0.018 |
| Credibility | 2.855 | 0.013 |

## Comparison with Study 1

To fully understand the effect of seeing only three elements when making a decision, it is relevant to compare with the data in Study 1's Hidden and Visible condition. This allows for a clear comparison to be made of the effect of amount of information on participants' decision-making.

Comparing the data from the 3-Seen and 3-Reveal with the Hidden condition revealed significant differences in participants' responses. A main effect of Profile condition was seen for the rent decision, $F(2, 154) = 4.27$, $p = .016$, $\eta^2 = .053$. Post-hoc Tukey HSD comparisons revealed a significant difference between the Hidden and the 3-Reveal condition ($M = 7.93$, $SD = 2.18$), $p = .011$, echoing Study 1. However, there was no differences in scores between the 3-Seen ($M = 7.34$, $SD = 2.58$) and 3-Reveal condition, or 3-Seen and Hidden condition. With respect to confidence ratings, no significant difference was found based on profile condition, F < 1, ns.

Considering credibility, differences were found based on profile condition, $F(2, 154) = 4.30$, $p = .015$, $\eta^2 = .053$. Post-hoc comparisons revealed a significant difference between the Hidden and 3-Reveal condition ($M = 75.25$, $SD = 13.18$), $p = .012$, and a marginally significant difference between the Hidden and 3-Seen condition ($M = 73.00$, $SD = 11.80$), $p = .092$.

For trust towards the host, a main effect of profile was found, $F(2, 154) = 7.09$, $p = .001$, $\eta^2 = .084$. Post-hoc comparisons revealed a significant difference between the Hidden

condition and both the 3-Seen ($M$ = 70.70, $SD$ = 11.37), $p$ = .023, and the 3-Reveal ($M$ = 73.77, $SD$ = 13.79), $p$ = .001.

A marginally significant effect of profile was also seen for sociability, $F(2, 154)$ = 2.73, $p$ = .069, $\eta^2$ = .034. Post-hoc comparisons revealed only a marginally significant difference between the Hidden condition and the 3-Reveal condition ($M$ = 67.68, $SD$ = 15.45), $p$ = .060.

Finally, an analysis comparing Study 2's conditions with Study 1's Visible condition was performed, assessing differences in user judgement when more information was available. This did not reveal any significant differences on any of the measured DVs, $p$s > .13.

Thus, when comparing with Study 1's Hidden condition it was found that the same positivity towards hosts was observers in users' data. But, when comparing to Study 1's Visible condition, no differences were found. This supports the prediction that users focus/use primarily on around 3 elements when making rental choices and judging hosts.

# S6 Study 3 Demographics and Supplementary Analyses

## Participant Demographics

| Ethnicity | | | | | | |
|---|---|---|---|---|---|---|
| **Asian** | **Black** | **Latino** | **White** | **Multi-racial** | **Other** | **N** |
| 18 | 16 | 12 | 139 | 0 | 4 | 189 |

| Sharing Economy - Memberships | | | | | |
|---|---|---|---|---|---|
| **None** | **1-2** | **3-4** | **5-10** | **>10** | **N** |
| 24 | 129 | 34 | 2 | 0 | 189 |

| Sharing Economy – Usage Length | | | | | |
|---|---|---|---|---|---|
| **<1 month** | **2-6 months** | **8-10 months** | **12 months** | **>3 years** | **N** |
| 32 | 38 | 27 | 52 | 40 | 189 |

| Sharing Economy - Usage Frequency | | | | | |
|---|---|---|---|---|---|
| **0-5** | **5-10** | **10-20** | **20-30** | **>30** | **N** |
| 85 | 55 | 26 | 6 | 17 | 189 |

| Sharing Economy - Satisfaction | | | | | | | | | |
|---|---|---|---|---|---|---|---|---|---|
| **1** | **3** | **4** | **5** | **6** | **7** | **8** | **9** | **10** | **N** |
| 10 | 1 | 6 | 25 | 17 | 38 | 44 | 29 | 19 | 189 |

| Sharing Economy – Sense of Belonging | | | | | | | | | |
|---|---|---|---|---|---|---|---|---|---|
| **1** | **3** | **4** | **5** | **6** | **7** | **8** | **9** | **10** | **N** |
| 20 | 8 | 9 | 14 | 25 | 20 | 36 | 32 | 16 | 189 |

## Manipulation Check

As in Study 2, preliminary analyses found no differences between any of the sub-conditions used in the study for any of the measured DVs, $F$s < 1, $p$s > .05.

Several independent-samples t-test were conducted to ensure that the 3-Wanted condition was not significantly different from the Study 2's 3-Seen condition. Considering each DV, no significant differences were found, $t$s ≤ 1, $p$s > .05, suggesting the conditions can be treated equally.

Finally, a gender analysis between the three conditions did not reveal any statistically significant differences on any of the measured DVs, $F$s ≤ 1, $p$s > .05.

## Comparison with Study 1

To understand if the triplets seen by participants in the three profile conditions had an effect on their ratings of hosts and decisions to rent, the data from Study 3' 3-Random and 3-Avoided was compared with the Hidden, Reveal, and Visible conditions from Study 1. This allows for a direct comparison of how the type and amount of TRI impacts users' decision-making.

***Random vs Hidden.*** Conducting independent-samples t-tests between the 3-Random and S1-Hidden condition for each relevant DV revealed significant differences in line with our predictions.

For rent decisions, a significant difference was found where users in the 3-Random ($M$ = 8.16, $SD$ = 2.34) condition rented on average more private rooms than those in the S1-

Hidden condition, $t(94) = 3.43$, $p = .001$, 95% CI [.71, 2.66], $d = 0.71$.

For confidence, no difference was found between the two conditions, $t(94) = 1.08$, $p = .283$, 95% CI [-2.34, 7.89].

For sociability ratings, a significant difference was found between the two conditions, where users in the 3-Random ($M = 69.77$, $SD = 12.54$) condition rated hosts higher on sociability than those in the S1-Hidden condition, $t(94) = 3.06$, $p = .003$, 95% CI [3.19, 14.94], $d = 0.62$.

For trustworthiness ratings, users in the 3-Random ($M = 73.96$, $SD = 12.76$) condition rated hosts significantly higher than those in the S1-Hidden condition, $t(94) = 3.63$, $p < .001$, 95% CI [4.78, 16.35], $d = 0.74$.

For credibility ratings, a significant difference was found between the two conditions. Users in the 3-Random ($M = 77.88$, $SD = 11.73$) condition rated hosts higher on credibility than those in the S1-Hidden condition, $t(94) = 3.86$, $p < .001$, 95% CI [5.09, 15.91], $d = 0.78$.

***Random vs Visible.*** When comparing the 3-Random condition data with that of Study 1's Visible condition, containing all elements of TRI, no significant differences are found on any of the measured DVs (all $t$s ≤ 1.36, $p$s > .176).

***Avoided vs Hidden.*** Comparing the 3-Avoided triplet profile data with the S1-Hidden data revealed the same pattern of results as with the 3-Random comparisons.

Users made significantly more rent decisions in the 3-Avoided condition ($M = 7.73$, $SD = 2.79$) than in the S1-Hidden condition, $t(105) = 2.36$, $p = .20$, 95% CI [.20, 2.31], $d = 0.48$. While, again, no difference in confidence ratings were uncovered, $t < 1$, ns.

For host ratings, a significant difference was found for sociability, trustworthiness, and credibility as a result of the profile condition. Users rated hosts higher on sociability in the 3-Avoided condition ($M = 70.30$, $SD = 15.43$) than in the S1-Hidden condition, $t(105) = 2.36$, $p = .003$, 95% CI [3.33, 15.87], $d = 0.60$. They also rated hosts higher on

trustworthiness in the 3-Avoided condition ($M = 73.94$, $SD = 16.07$), $t(105) = 3.31$, $p = .001$, 95% CI [4.22, 16.86], $d = 0.66$. And, rated hosts higher on perceived credibility (3-Avoided, $M = 75.93$, $SD = 16.02$, $t(105) = 2.74$, $p = .007$, 95% CI [2.36, 14.74], $d = 0.55$.

*Avoid vs Visible.* As expected, when comparing the 3-Avoided condition responses with those of the S1-Visible, no significant differences were uncovered (all $t$s > .80, $p$s > .423). This suggests that three cues, even those users tend to not selected when given the opportunity, result in the same increased ratings towards hosts and decisions to rent.

*Random vs Reveal S1 and S2.* To ensure that the act of selecting information did not have an additional impact on user judgements beyond simply seeing three elements, comparison analyses were conducted between the 3-Random condition and the S1-Reveal condition data. Once more, the results did not indicate any significant differences in user responses, on any measure, all $t$s > .642, $p$s > .523; the same pattern was observed for the S2-3-Reveal condition comparison, $t$s > 1.114, $p$s > .268.

*Avoided vs Reveal S1 and S2.* Finally, the data from the 3-Avoided condition was similarly compared to that of Study 1's Reveal condition. This also did not produce any significant differences in users judgements between seeing a triplet with elements users tend to avoid selecting when making their rental decisions with those they select themselves, on any measure (S1-Reveal, all $t$s > .620, $p$s > .537; S2-3-Reveal, all $t$s > .937, $p$s > .350).

**Conclusion**

Thus, even when compared with the previous data where users had minimal (S1-Hidden) or full data (S1-Visible), the identical pattern of results was obtained as in Study 1 and 2. Overall, the data strongly supports the prediction that three elements of TRI are sufficient to affect user judgement relating to hosts and rental decision.